\documentclass[iop,apj,superscriptaddress]{emulateapj}
\usepackage{amsmath}
\usepackage{graphicx}
\usepackage{amssymb}
\usepackage{bm}
\usepackage{braket}
\usepackage{hyperref}
\usepackage{natbib}
\shorttitle{Current singularity in line-tied plasma}
\shortauthors{Zhou, Huang, Qin, \& Bhattacharjee}
\begin{document}

\title{Constructing current singularity in a 3D line-tied plasma}

\author{Yao Zhou, Yi-Min Huang, Hong Qin\altaffilmark{1}, and A. Bhattacharjee}
\affiliation{Department of Astrophysical Sciences and Plasma Physics Laboratory, Princeton University, Princeton, NJ 08543, USA}
\altaffiltext{1}{Department of Modern Physics, University of Science and Technology of China, Hefei, Anhui 230026, China}

%\author{Yao Zhou}
%\affiliation{Department of Astrophysical Sciences and Plasma Physics Laboratory, Princeton University, Princeton, NJ 08543, USA}
%\author{Yi-Min Huang}
%\affiliation{Department of Astrophysical Sciences and Plasma Physics Laboratory, Princeton University, Princeton, NJ 08543, USA}
%
%\author{Hong Qin}
%\affiliation{Department of Astrophysical Sciences and Plasma Physics Laboratory, Princeton University, Princeton, NJ 08543, USA}
%\affiliation{Department of Modern Physics, University of Science and Technology of China, Hefei, Anhui 230026, China}
%
%\author{A. Bhattacharjee}
%\affiliation{Department of Astrophysical Sciences and Plasma Physics Laboratory, Princeton University, Princeton, NJ 08543, USA}



\begin{abstract}
We revisit Parker's conjecture of current singularity formation in 3D line-tied plasmas using a recently developed numerical method, variational integration for ideal magnetohydrodynamics in Lagrangian labeling. With the frozen-in equation built-in, the method is free of artificial reconnection, and hence it is arguably an optimal tool for studying current singularity formation. Using this method, the formation of current singularity has previously been confirmed in the Hahm--Kulsrud--Taylor problem in 2D. In this paper, we extend this problem to 3D line-tied geometry. The linear solution, which is singular in 2D, is found to be smooth for arbitrary system length. However, with finite amplitude, the linear solution can become pathological when the system is sufficiently long. The nonlinear solutions turn out to be smooth for short systems. Nonetheless, the scaling of peak current density versus system length suggests that the nonlinear solution may become singular at finite length. With the results in hand, we can neither confirm nor rule out this possibility conclusively, since we cannot obtain solutions with system length near the extrapolated critical value.

\end{abstract}

\keywords{magnetohydrodynamics (MHD) --- magnetic fields 
 --- Sun: corona}
%\maketitle


\section{Introduction.}
A long-standing problem in solar physics is why the solar corona, a nearly perfectly conducting plasma, where the Lundquist number $S$ can be as high as $10^{14}$, has an anomalously high temperature that conventional Ohmic heating cannot explain. Decades ago, \citet{Parker1972} proposed that the convective motions in the photosphere will tend to induce current singularities in the corona, and the subsequent magnetic reconnection events can account for substantial heating. This conjecture has remained controversial to this day \citep{Rosner1982,Parker1983,Parker1994,Tsinganos1984,VanBallegooijen1985,VanBallegooijen1988,Zweibel1985,Zweibel1987,Antiochos1987,Longcope1994b,Ng1998,Longbottom1998,Bogoyavlenskij2000,Craig2005,Low2006,Low2010,Wilmot-Smith2009,Wilmot-Smith2009b,Janse2010,Rappazzo2013,Craig2014,Pontin2015,Candelaresi2015,Pontin2016}.

This controversy fits into the larger context of current singularity formation, which is also a problem of interest in toroidal fusion plasmas \citep{Grad1967,Rosenbluth1973,Hahm1985,Loizu2015b}. However, the solar corona, where magnetic field lines are anchored in the photosphere, is often modeled with the so-called line-tied geometry. This is a crucial difference from toroidal fusion plasmas where closed field lines can exist. For clarification, in this article, we refer to the problem of whether current singularities can emerge in 3D line-tied geometry as \textit{the Parker problem}.

Although this problem is inherently dynamical, it is usually treated by examining magnetostatic equilibria for simplicity, as \citet{Parker1972} originally did. The justification is, if the final equilibrium that an initially smooth magnetic field relaxes to contains current singularities, they must have formed during the relaxation. A key assumption here is that the plasma is perfectly conducting, so the equilibrium needs to preserve the magnetic topology of the initial field. Analytically, this topological constraint is difficult to explicitly attach to the magnetostatic equilibrium equation. Numerically, most standard methods for ideal magnetohydrodynamics (MHD) are susceptible to artificial field line reconnection in the presence of (nearly) singular current densities. Either way, to enforce this topological constraint is a major challenge for studying the Parker problem. 

Fortunately, one can overcome this difficulty by adopting Lagrangian labeling, where the frozen-in equation is built into the equilibrium equation, instead of the commonly used Eulerian labeling. \citet{Zweibel1987} first noticed that this makes the mathematical formulation of the Parker problem explicit and well-posed. Moreover, not solving the frozen-in equation numerically avoids the accompanying error and resultant artificial reconnection. A Lagrangian relaxation scheme with this feature has been developed using conventional finite difference \citep{Craig1986}, and extensively used to study the Parker problem  \citep{Longbottom1998,Craig2005,Wilmot-Smith2009,Wilmot-Smith2009b,Craig2014}. 
\citet{Pontin2009} later found that its current density output can violate charge conservation, and mimetic discretization has been applied to fix it \citep{Candelaresi2014}. 

Recently, a variational integrator for ideal MHD in Lagrangian labeling has been developed by \citet{Zhou2014} using discrete exterior calculus \citep{Desbrun2005}. Derived in a geometric and field-theoretic manner, it naturally preserves many of the conservation laws of ideal MHD, including charge conservation. It is arguably an optimal tool for studying current singularity formation.

\citet{Zhou2016} have used this method to study the Hahm--Kulsrud--Taylor (HKT) problem \citep{Hahm1985}, a fundamental prototype problem for current singularity formation in 2D, where a plasma in a sheared magnetic field is subject to boundary forcing. The formation of current singularity is conclusively confirmed via convergence study, and its signature is also identified in other 2D cases with more complex topologies, such as the coalescence instability of magnetic islands \citep{Longcope1993}.

In this paper, we extend the HKT problem to 3D line-tied geometry. \citet{Zweibel1987} showed that the linear solution, which is singular in 2D, should become smooth. This prediction is confirmed by our numerical results. However, we also find that given finite amplitude, the linear solution can be pathological when the system is sufficiently long. We speculate that this finite-amplitude pathology may trigger a finite-length singularity in the nonlinear solution.

We perform a convergence study on the nonlinear solutions for varying system length $L$.
For short systems, the nonlinear solutions converge to smooth ones. The peak current density approximately scales with $(L_n-L)^{-1}$, suggesting that the solution may become singular above a finite length $L_n$. However, the solutions for longer systems inherently involve strongly sheared motions, which {lead to severe mesh distortions that terminate our numerical simulations}. As a result, we cannot obtain solutions for systems with lengths close to $L_n$, and hence cannot conclude whether such a finite-length singularity does exist. Nonetheless, our results are suggestive that current singularity may well survive in this line-tied system, in accordance with the arguments in \citet{Ng1998}.

This paper is organized as follows. In Sec.\,\ref{ParkerP} we formulate the Parker problem in Lagrangian labeling, specify the setup in line-tied geometry, and introduce the conventions of reduced MHD. Our numerical method is illustrated in Sec.\,\ref{numerical}. In Sec.\,\ref{HKT} we review the conclusions from the HKT problem in 2D, and then present our results, both linear and nonlinear, in 3D line-tied geometry. Discussions follow in Sec.\,\ref{discussion}.

\section{the Parker problem}\label{ParkerP}
\citet{Parker1972} originally considered a perfectly conducting plasma, magnetized by a uniform field $\mathbf{B}=\hat z$ threaded between two planes at $z=0,L$, which are often referred to as the footpoints. The footpoints are then subject to random motions such that the magnetic field becomes nonuniform. He argued that in general, there exists no smooth equilibrium for the system to relax to, and therefore current singularities must form. This conjecture is based on perturbative analysis of the magnetostatic equilibrium equation,
\begin{equation}
(\nabla\times\mathbf{B})\times\mathbf{B}=\nabla p,\label{JcrossB}
\end{equation}
where $p$ is the pressure. Many of the subsequent works on the Parker problem are performed on this equation as well \citep{Rosner1982,Parker1983,Tsinganos1984,VanBallegooijen1985,Antiochos1987,Bogoyavlenskij2000,Low2006,Low2010,Janse2010}.

A caveat of this approach is that Eq.\,(\ref{JcrossB}) is usually underdetermined. That is, a given set of boundary conditions may allow for more than one solution to this equation, and additional information is needed to identify a specific one. {Often, it is prescribed as part of the equilibrium solution. For example, in 2D Eq.\,(\ref{JcrossB}) reduces to the Grad--Shafranov equation \citep{Grad1967}, where the pressure and guide field profiles are prescribed to the equilibrium.} For the Parker problem, the information is the very constraint to preserve the initial magnetic topology. The implication is that identifying singular solutions to Eq.\,(\ref{JcrossB}) does not necessarily prove Parker's conjecture, since these solutions may not satisfy this topological constraint.

However, this topological constraint is mathematically challenging to explicitly attach to Eq.\,(\ref{JcrossB}) and its solutions \citep{Janse2010,Low2010}. Nonetheless, it can be naturally enforced if one adopts Lagrangian labeling for ideal MHD, instead of Eulerian labeling, which is used in Eq.\,(\ref{JcrossB}), as first noticed by \citet{Zweibel1987}.

\subsection{Lagrangian labeling}
In Lagrangian labeling, the motion of the fluid elements is traced in terms of a continuous mapping from the initial position $\mathbf{x}_0$ to the current position $\mathbf{x}(\mathbf{x}_0,t)$, which is also referred to as the fluid configuration. In this formulation, the advection (continuity, adiabatic, and frozen-in) equations are \citep{Newcomb1962}
\begin{subequations}\label{advection}
\begin{align}
\rho\,\mathrm{d}^3{x}=\rho_0\,\mathrm{d}^3x_0&\Rightarrow \rho=\rho_0/J,\label{continuity}\\
p/\rho^\gamma=p_0/\rho_0^\gamma&\Rightarrow p=p_0/J^\gamma,\label{adiabatic}\\
B_i\,\mathrm{d}S_i=B_{0i}\,\mathrm{d}S_{0i}&\Rightarrow B_i=x_{ij}B_{0j}/J.\label{frozenin}
\end{align}
\end{subequations}
{Here Einstein summation is implied for the subscripts $i,j\in \{1,2,3\}$, which denote the components of vectors. They should be distinguished from the subscript $0$ that denotes values at $t=0$: $\rho_0=\rho(\mathbf{x}_0,0)$ is the initial mass density, and the same goes for $p_0$ and $\mathbf{B}_0$.
In addition, $x_{ij}=\partial x_i/\partial x_{0j}$ and $J=\det(x_{ij})$ are the elements and the determinant of the Jacobian matrix, respectively, while $\gamma$ is the adiabatic index.}
%$\rho_0=\rho(\mathbf{x}_0,0)$, $p_0=p(\mathbf{x}_0,0)$, $\mathbf{B}_0=\mathbf{B}(\mathbf{x}_0,0)$
%are the initial mass density, pressure, and magnetic field respectively. 

Eq.\,\eqref{advection} reflects the fact that in ideal MHD, mass, entropy, and magnetic flux are advected by the motion of the fluid elements. They are built into the ideal MHD Lagrangian and the subsequent Euler-Lagrange equation \citep{Newcomb1962},
\begin{align}
&\rho_0\ddot{x}_i-B_{0j}\frac{\partial}{\partial x_{0j}}\left(\frac{x_{ik}B_{0k}}{J}\right)\nonumber\\
&+C_{ij}\frac{\partial }{\partial x_{0j}}\left(\frac{p_0}{J^\gamma}+\frac{x_{kl}x_{km}B_{0l}B_{0m}}{2J^2}\right)=0.\label{momentum3}
\end{align}
{Here $C_{ij}=\partial J/\partial x_{ij}$ is the cofactor of $x_{ij}$ in $J$.} Eq.\,\eqref{momentum3} is the momentum equation, the only ideal MHD equation in Lagrangian labeling. 

Without time dependence, Eq.\,(\ref{momentum3}) becomes an equilibrium equation. Its solutions will satisfy not only Eq.\,(\ref{JcrossB}) but automatically the topological constraint implied in the Parker problem, since the initial field configuration $\mathbf{B}_0$ is built into the equation. In contrast, not all solutions to Eq.\,(\ref{JcrossB}) can necessarily be mapped from given initial conditions. 

Thus, the equilibrium equation in Lagrangian labeling offers a more natural and mathematically explicit description for the Parker problem, which simply becomes whether there exist singular solutions to this equation, given certain smooth initial and boundary conditions. If the initial field $\mathbf{B}_0$ is smooth, any singularity in the equilibrium field $\mathbf{B}$ should trace back to that in the fluid mapping $\mathbf{x}(\mathbf{x}_0)$. In Sec.\,\ref{geometry}, we will specify the initial and boundary conditions for the Parker problem.

\subsection{3D line-tied geometry}\label{geometry}
Parker's original model can be characterized with a uniform initial field $\mathbf{B}_0=\hat z$ and prescribed smooth footpoint motion $\mathbf{x}_{\perp}(\mathbf{x}_{0\perp})$ at ${z_0=0,L}$ while $z|_{z_0=0,L}=z_0$. The subscript $\perp$ denotes the in-plane components ($x,y$). Certain classes of footpoint motion were considered in \citet{VanBallegooijen1988,Mikic1989,Longbottom1998,Ng1998,Craig2005}, which are referred to as braiding experiments in the recent review by \citet{Wilmot-Smith2015}.

An alternative is to consider a nonuniform $\mathbf{B}_0$, referred to as initially braided field in \citet{Wilmot-Smith2015}, with no-slip footpoints ($\mathbf{x}=\mathbf{x}_{0}$ at $z_0=0,L$). Note that to remain relevant to Parker's original model, the nonuniform $\mathbf{B}_0$ must be realizable from Parker's uniform field via smooth footpoint motion. Examples include the coalescence instability \citep{Longcope1994,Longcope1994b} and the threaded X-point \citep{Craig2014}, but exclude those with magnetic nulls \citep{Pontin2005,Craig2014b}.

We adopt the latter approach for its two advantages. One is reduced computational complexity. More importantly, these initially braided fields are usually extended from 2D cases that are susceptible to current singularity formation \citep{Longcope1993,Craig2005b}. Unlike Parker's original setup, this allows one to focus on the effect of 3D line-tied geometry on current singularity formation. In Sec.\,\ref{HKT}, we will extend the HKT problem \citep{Hahm1985}, where current singularity formation is confirmed in 2D \citep{Zhou2016}, to 3D line-tied geometry. 

{It is worth noting that \citet{Parker1972} did not specify the in-plane boundary conditions in his original setup, but assumed the system to be infinitely large. In numerical studies, in-plane boundaries are most commonly set to be periodic \citep[e.g.,][]{Longcope1994,Longbottom1998,Rappazzo2013}. Occasionally, other boundary conditions such as line-tying have also been used in the in-plane directions \citep{Craig2014,Candelaresi2015}. In the HKT problem that we shall study, the boundaries in one of the in-plane directions are perfectly conducting walls. The justifications and implications will be discussed in Sec.\,\ref{HKT}.}

\subsection{Reduced MHD}\label{RMHD}
Reduced MHD \citep[RMHD,][]{Strauss1976} is a reduction of MHD in the strong guide field limit that is often used to model the solar corona. \citet{VanBallegooijen1985} first used what essentially is RMHD to study the Parker problem, {as followed by many others \citep[e.g.,][]{Longcope1994b,Ng1998,Rappazzo2013}.}

In Eulerian labeling, RMHD approximations include uniform guide field ($B_{z}=1$), removal of $z$ dynamics (${v}_z=0$), and incompressibility ($\nabla\cdot\mathbf{v}=0$). The equilibrium equation becomes
\begin{equation}
\mathbf{B}\cdot\nabla j_z=0,\label{reduced}
\end{equation}
which is obtained from the $z$ component of the curl of Eq.\,(\ref{JcrossB}). Here $\mathbf j = \nabla\times\mathbf B$ is the current density.

Physically, Eq.\,(\ref{reduced}) means that $j_z$ is constant along a field line. In RMHD, every field line is threaded through all $z$. Therefore, the implication for the Parker problem is, if an equilibrium solution yields a current singularity, it must penetrate into the line-tied boundaries. Note that this is a very strong condition that applies to all solutions of Eq.\,(\ref{reduced}), topologically constrained or not.
 
 Translated into Lagrangian labeling, RMHD approximations become $B_{0z}=1$, $z=z_0$, and $J=1$. Following Eq.\,(\ref{frozenin}), the in-plane field then reads
  \begin{equation}\label{Bperp}
\mathbf{B}_{\perp}=\frac{\partial\mathbf{x}_{\perp}}{\partial\mathbf{x}_{0\perp}}\cdot\mathbf{B}_{0\perp}+\frac{\partial\mathbf{x}_{\perp}}{\partial z_{0}}.
\end{equation}
The first term on the RHS results from in-plane motion, while the second term is the projection of the tilted guide field that shows up only in 3D. 

At the line-tied boundaries ($z_0=0,L$), where $\mathbf{x}_{\perp}=\mathbf{x}_{0\perp}$, the $z$ component of the (Eulerian) curl of Eq.\,(\ref{Bperp}) reads
\begin{equation}\label{jz}
j_{z}\hat z=j_{0z}\hat z+\nabla_{\perp}\times\frac{\partial\mathbf{x}_{\perp}}{\partial z_{0}}.
\end{equation}
Here $j_{0z}$ is the initial condition that has to be smooth. That is, for $j_z$ to be (nearly) singular at the footpoints, $(\partial\mathbf{x}_{\perp}/\partial z_{0})|_{z_0=0,L}$ must be (nearly) singularly sheared (note that this is compatible with the line-tied boundary condition). Therefore, we assert that strongly sheared motion is an inherent feature of the Parker problem.

Throughout the rest of the paper, RMHD approximations are adopted unless otherwise noted. {One motivation is, there exist significant analytical studies in RMHD that can be compared with \citep{VanBallegooijen1985,Longcope1994b,Ng1998}. Another is due to numerical concerns that will be explained in Sec.\,\ref{nonlinear}.}

{We expect that in 3D line-tied geometry, if a current singularity can emerge in RMHD, it will likely survive in full MHD. In full MHD, Eq.\,\eqref{reduced} does not hold, so the stringent condition that a current singularity must penetrate into the line-tied boundaries might be relaxed. Also, RHMD precludes the boundary layers close to the footpoints that are identified in full MHD analysis \citep{Zweibel1985,Zweibel1987,Scheper1999}, which makes current singularity formation relatively more difficult in RMHD.}

{All that being said, we acknowledge that the RMHD framework we adopt is a rather strong simplification of full MHD, and our results and conclusions may be subject to changes in the latter.}

\section{numerical method}\label{numerical}
Numerically, many have used Eulerian methods for ideal MHD to study the Parker problem \citep{Mikic1989,Longcope1994,Ng1998,Rappazzo2013}. These simulations all end up encountering artificial reconnection, and topologically constrained equilibrium solutions cannot be obtained.

In contrast, Lagrangian methods that solve Eq.\,(\ref{momentum3}) with moving meshes avoid solving the frozen-in equation and the consequent artificial reconnection. For example, a Lagrangian relaxation scheme \citep{Craig1986} has been extensively used to study the Parker problem  \citep{Longbottom1998,Craig2005,Wilmot-Smith2009,Wilmot-Smith2009b,Craig2014}. In this method, the inertia (first term) in Eq.\,(\ref{momentum3}) is replaced by frictional damping, which has been argued to cause unphysical artifacts by \cite{Low2013}. Also, \citet{Pontin2009} showed that the spatial discretization using conventional finite difference can violate charge conservation ($\nabla\cdot\mathbf{j}=0$). Both of these issues have been fixed in \citet{Candelaresi2014}: the former by retaining the inertia during the frictional relaxation, and the latter with mimetic discretization.

The numerical method we use is a recently developed variational integrator for ideal MHD \citep{Zhou2014}. It is obtained by discretizing the Lagrangian for ideal MHD in Lagrangian labeling \citep{Newcomb1962} on a moving unstructured mesh. Using discrete exterior calculus \citep{Desbrun2005}, the momentum equation (\ref{momentum3}) is spatially discretized into a conservative many-body form $M_i\ddot{\mathbf{x}}_i=-\partial V/\partial \mathbf{x}_i$, where $M_i$ and $\mathbf{x}_i$ are the mass and position of the $i$th vertex, respectively, and $V$ is a spatially discretized potential (internal plus magnetic) energy. When the system is integrated in time, friction may be introduced to dynamically relax it to an equilibrium with minimal $V$ by dissipating the momentum. Friction does not affect the advection equations \eqref{advection} since they are built into the equilibrium equation.

Compared with similar methods \citep{Craig1986,Candelaresi2014}, our method exactly preserves many conservation laws. {For example, charge conservation is guaranteed due to a discrete Stokes' theorem featured by discrete exterior calculus  \citep{Desbrun2005}, which is more general than the mimetic methods adopted by \citet{Candelaresi2014}}. Our discrete force is conservative, which means the equilibrium solution minimizes a discrete potential energy. Constructed on unstructured meshes, the method allows resolution to be devoted to where it is most needed, such as the vicinity of a potential current singularity. {Readers interested in the numerical method are referred to \citet{Zhou2014} and the references therein for further details.}

An Achilles' heel of our numerical method, and others that solve Eq.\,(\ref{momentum3}) with moving meshes, is its vulnerability to {severe mesh distortion} due to strong shear flow. Unfortunately, as discussed in Sec.\,\ref{RMHD}, strongly sheared motion is an inherent feature of the Parker problem, posing a formidable challenge for our numerical endeavor at the very outset.

%The scheme inherits built-in advection equations from the continuous formulation, and thus avoids the error and dissipation associated with solving them. It has been shown that the scheme can handle prescribed singular current sheets without suffering from artificial field line reconnection. Such capability of preserving the magnetic topology makes it an optimal tool for studying current sheet formation. 

\section{The HKT problem} \label{HKT}
The HKT problem was originally proposed by Taylor and studied by \citet{Hahm1985} in the context of studying forced magnetic reconnection induced by resonant perturbation on a rational surface. It is a fundamental prototype problem considering how a 2D incompressible plasma magnetized by a sheared equilibrium field $B_{0y} = x_0$ responds to external perturbations. Specifically, the perfectly conducting boundaries at $x_0=\pm a$ are deformed sinusoidally into the shapes that $x(\pm a, y_0)=\pm[a-\delta\cos ky(\pm a, y_0)$]. 

\citet{Zweibel1987} first connected this problem to the Parker problem, since the sheared initial field is easily realizable from Parker's uniform field via sheared footpoint motion. In 2D, Their linear equilibrium solution in Lagrangian labeling reads
\begin{align}\label{HKS}
\chi=-\frac{\delta a\sinh kx_0\sin ky_0}{k|x_0|\sinh ka},
\end{align}
where $\chi$ is the stream function for the linear displacement $\bm\xi = \nabla \chi\times \hat z $. %Note that the linear equilibrium equation in Lagrangian labeling {can equivalently be obtained by  linearizing Eq.\,\eqref{reduced} while using the frozen-in constraint \eqref{frozenin} to express the perturbed quantities in terms of $\chi$}

%is simply $\mathbf{F}(\bm\xi )=0$,. Here $\mathbf{F}(\bm\xi )$ is the standard ideal MHD force operator \citep{Schnack2009linearized}. 

The linear solution (\ref{HKS}) yields a perturbed magnetic field $\delta B_y=\text{sgn}(x_0)ka\delta \cosh kx_0\cos ky_0/\sinh ka$, which contains a current singularity at the neutral line $x_0=0$. This singularity results from the one in $\partial\xi_x/\partial{x_0}$, but such a normal discontinuity in the displacement is not physically permissible (see Fig.\,\ref{pathology} and relevant discussion in Sec.\,\ref{finite}). The failure at the neutral line is expected from the linear solution since the linear assumption breaks down there. 

It has remained unclear whether the nonlinear solution to this problem is singular, until \citet{Zhou2016} used the numerical method described in Sec.\,\ref{numerical} to confirm it.
It is found that the equilibrium fluid mapping normal to the neutral line at $y_0=0$, namely $x(x_0,0)$, converges to a quadratic power law $x\sim  x_0^2$. Due to incompressibility, $(\partial y/\partial y_0)|_{y_0=0}\sim  x_0^{-1}$ diverges at $x_0=0$. With the initial field $B_{0y} = x_0$ substituted into Eq.\,(\ref{Bperp}), this mapping leads to an equilibrium field $B_y(x,0)\sim \text{sgn}(x)$ that is discontinuous at the neutral line. Here sgn stands for the sign fuction. 

Physically, this means the fluid element at the origin $(0,0)$ is infinitely compressed normally towards, while infinitely stretched tangentially along, the neutral line. In \citet{Zhou2017b}, exactly the same signature of current singularity is also identified in other 2D cases with more complex topologies, such as the coalescence instability of magnetic islands \citep{Longcope1993}. 

{Note that the coalescence instability is an internal ideal instability modeled in doubly periodic geometry. In contrast, the HKT problem is driven by external perturbations on the perfectly conducting boundaries. The fact that the exact same signature of current singularity emerges in both systems demonstrates the generality of this mechanism for current singularity formation in 2D, squashing a sheared magnetic field, regardless of different drives or boundary conditions.} 

{Furthermore, one can consider the HKT problem as a simplest prototype problem examining how a sheared magnetic field responds to squashing, which is modeled by the deformed perfectly conducting walls. In reality, squashing can be driven by instabilities such as the coalescence instability. }

{In fact, Parker extensively discussed (what essentially is) the HKT problem, which he termed as \textit{compression of a primitive (force-free) field}, in his monograph on the Parker problem \citep{Parker1994}. The bottom line is, with perfectly conducting boundaries, the HKT problem is no less relevant to the Parker problem than others with periodic in-plane boundaries.}

Then, the question becomes whether squashing a sheared magnetic field works in 3D line-tied geometry. We can learn from Eq.\,(\ref{Bperp}) that squashing is 2D in-plane motion that only contributes to the first term on the RHS. At the footpoints, where in-plane motion is absent and Eq.\,(\ref{jz}) holds, squashing is not allowed anymore. Hence, we expect 3D line-tied geometry to have a smoothing effect on the 2D current singularity. Yet we need to find out whether it eliminates the singularity entirely.

\subsection{Linear results}\label{linearR}
For the HKT problem in 2D, the singularity in the linear solution appears to be very suggestive for that in the nonlinear solution. Naturally, when extending the problem to 3D line-tied geometry, we consider the linear solution first.

In 3D line-tied geometry, we modulate the linear displacement on the perfectly conducting boundaries at $x_0 = \pm a$ into the form of $\xi_x(\pm a, y_0,z_0) = \mp \delta \cos ky_0\sin(\pi z_0/L)$. The perturbations vanish at the footpoints ($z_0 = 0, L$), which is consistent with the line-tied (no-slip) boundary condition. Accounting for the initial field $B_{0y}=x_0$, adopting RMHD approximations ($B_{0z}=1$ and $\bm\xi = \nabla \chi\times \hat z $) and Fourier dependence $\chi=\bar\chi(x_0,z_0)\exp{iky_0}$, the linear equilibrium equation becomes  
\begin{align}\label{linear}
(ikx_0 + \partial_{z_0})(\partial_{x_0}^2-k^2)(ikx_0 + \partial_{z_0})\bar\chi = 0.
\end{align}
When $\partial_{z_0}=0$, the 2D solution (\ref{HKS}) can be recovered. 

\begin{figure}[h]
\includegraphics[scale=0.45]{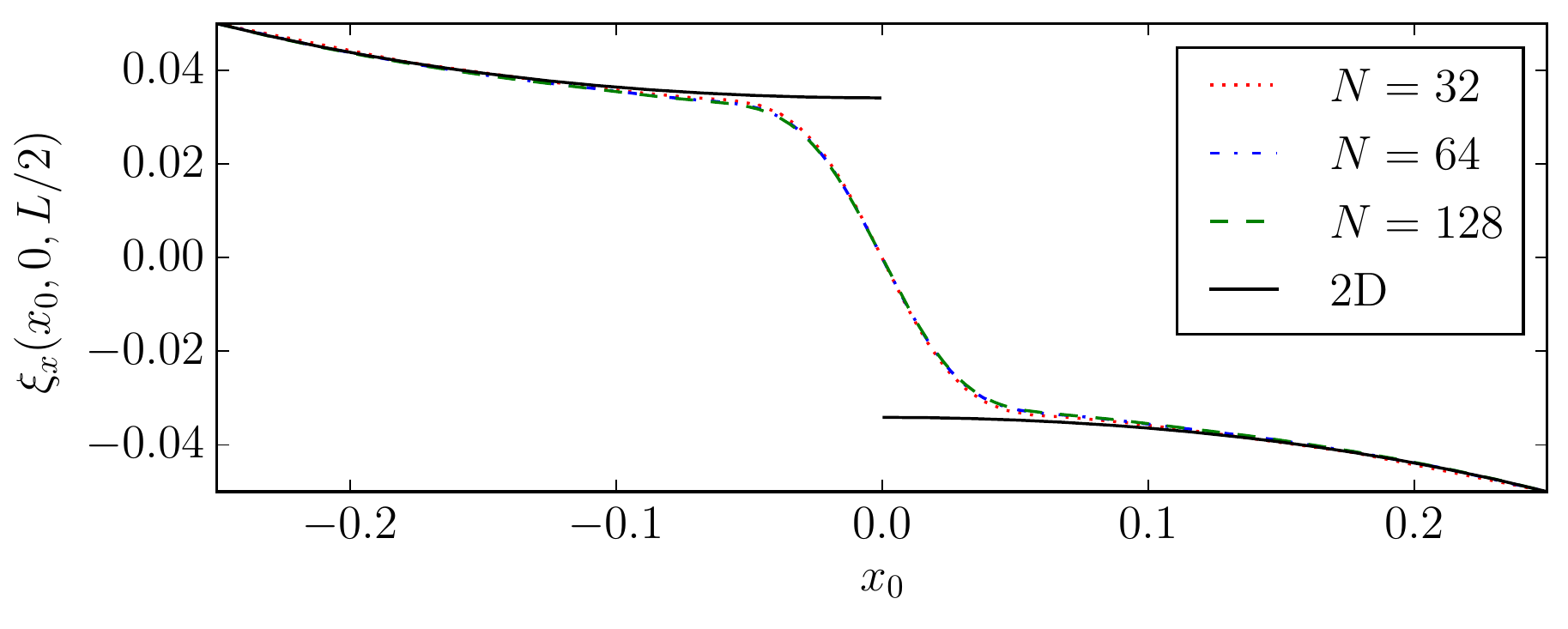}
\caption{\label{convergence}Numerical solutions of $\xi_x(x_0,0,L/2)$ for $L=32$ with $N=32$, $64$, and $128$ converge to a smooth one. The solid line shows the 2D solution (\ref{HKS}).}
\end{figure}

Eq.\,(\ref{linear}) is solved numerically using second-order centered finite difference, with boundary conditions $\bar\chi|_{x_0=\pm a}=\pm i(\delta/k)\sin(\pi z_0/L)$ and $\bar\chi|_{z_0=0,L}=0$. The parameters used are $a=0.25$, $k=2\pi$, $\delta=0.05$, with varying $L$ and resolution $N\times NL/8$. For a given $L$, the numerical solutions are found to converge to a smooth one. In Fig.\,\ref{convergence}, $\xi_x(x_0,0,L/2)$ obtained with different resolutions for $L=32$ are shown to converge.

\begin{figure}[h]
\includegraphics[scale=0.45]{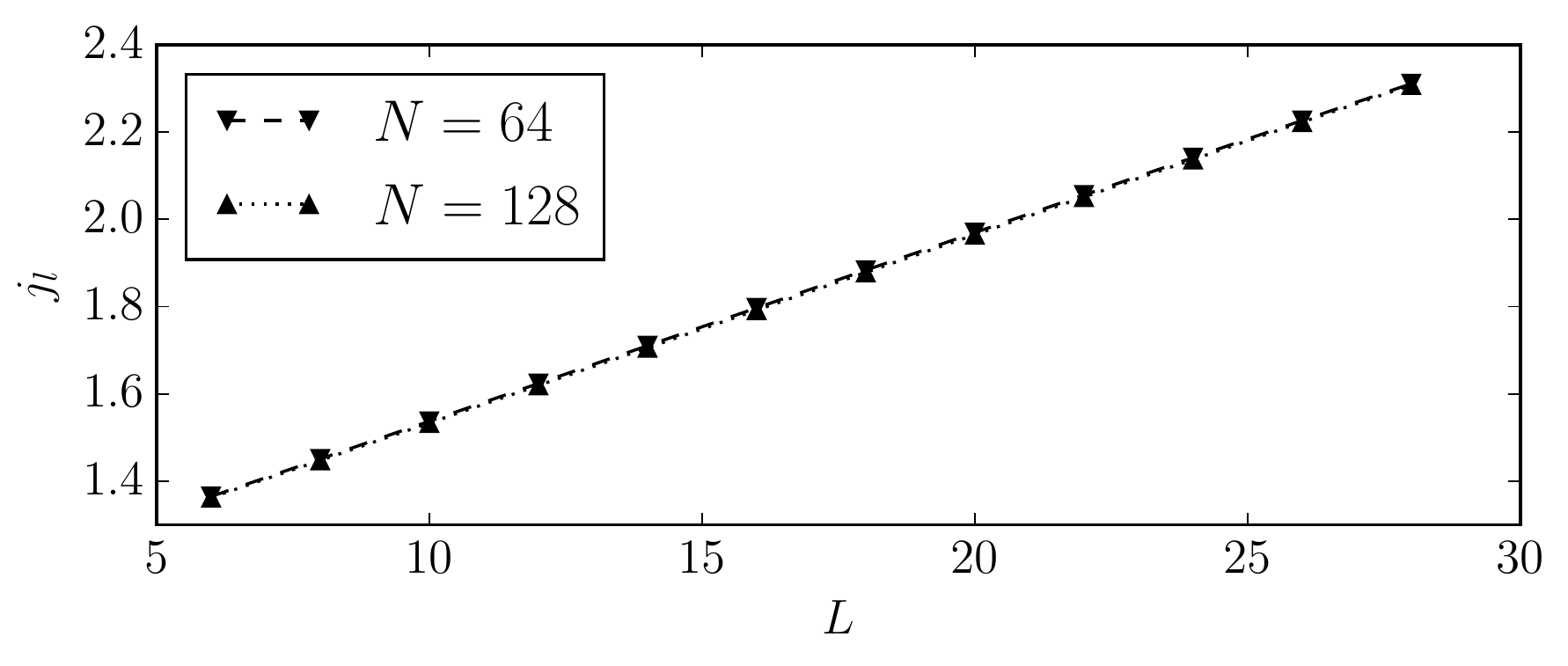}
\caption{\label{deltaj}Maximum of the linearly calculated current density $j_l$ vs. system length $L$ for $N=64$ and $128$. $j_l$ increases linearly with $L$.}
\end{figure}

We also find that with increasing $L$, $\xi_x(x_0,0,L/2)$ approaches the 2D solution with discontinuity (the solid line in Fig.\,\ref{convergence}) as its gradient at $x_0=0$ steepens. Accordingly, the maximum of the linearly calculated current density $j_{0z}+\delta j_z$, where $j_{0z}=1$ and $\delta j_z$ is the perturbed current density, is shown to increase linearly with $L$ in Fig.\,\ref{deltaj} (labeled $j_l$). This suggests that the linear solution is smooth for arbitrary $L$, only becoming singular when $L=\infty$. These results are consistent with the 3D linear analysis by \citet{Zweibel1987}.

So far, all the calculations have been strictly linear, assuming the amplitude of the perturbation $\delta$ to be infinitesimal. The linear solutions, be they $\xi_x$ or $\delta j_z$, are proportional to $\delta$. The magnitude of $\delta$ has no physical impact in this context.

The finite amplitude of the perturbation must be accounted for in the fully nonlinear study. Before that, we further exploit the linear solutions by considering the effect of finite amplitude on them in Sec.\,\ref{finite}.

\subsection{Finite-amplitude pathology}\label{finite}

%Now consider fluid mapping $\mathbf x =\mathbf x_0+\bm\xi$, where $\bm\xi$ is the linear equilibrium displacement with finite amplitude $\delta$. $\mathbf x$ is by no means a nonlinear equilibrium solution. That is, it does not satisfy the nonlinear equilibrium equation in Lagrangian labeling [Eq.\,(\ref{momentum3}) without time dependence]. Nonetheless, one can still calculate ``nonlinearly'' the magnetic field it maps into, using Eq.\,(\ref{frozenin}), and the current density $j_z$ thereafter, which we refer to as the finite-amplitude current density in this paper.

%We perform such calculation using the linear solutions obtained (with $\delta=0.05$) in Sec.\,\ref{linearR}, and notice that the finite-amplitude current density peaks at ${(0,0,L/2)}$. As shown in Fig.\,\ref{deltaj}, the maximum $j_{f}\sim(L_f-L)^{-2}$, which diverges at a critical length $L_f$. The value $L_f\approx28.96$ (using solutions with $N=128$ for fitting) depends on the specific parameters we obtain the linear solutions with, $\delta$ in particular.

It is mentioned above that for the HKT problem in 2D, the discontinuity in the linear solution (\ref{HKS}) is not physically permissible. In this section, we show that given finite amplitude, the linear solution in 3D line-tied geometry exhibits similar pathology when the system is sufficiently long.

\begin{figure}[h]
\includegraphics[scale=0.45]{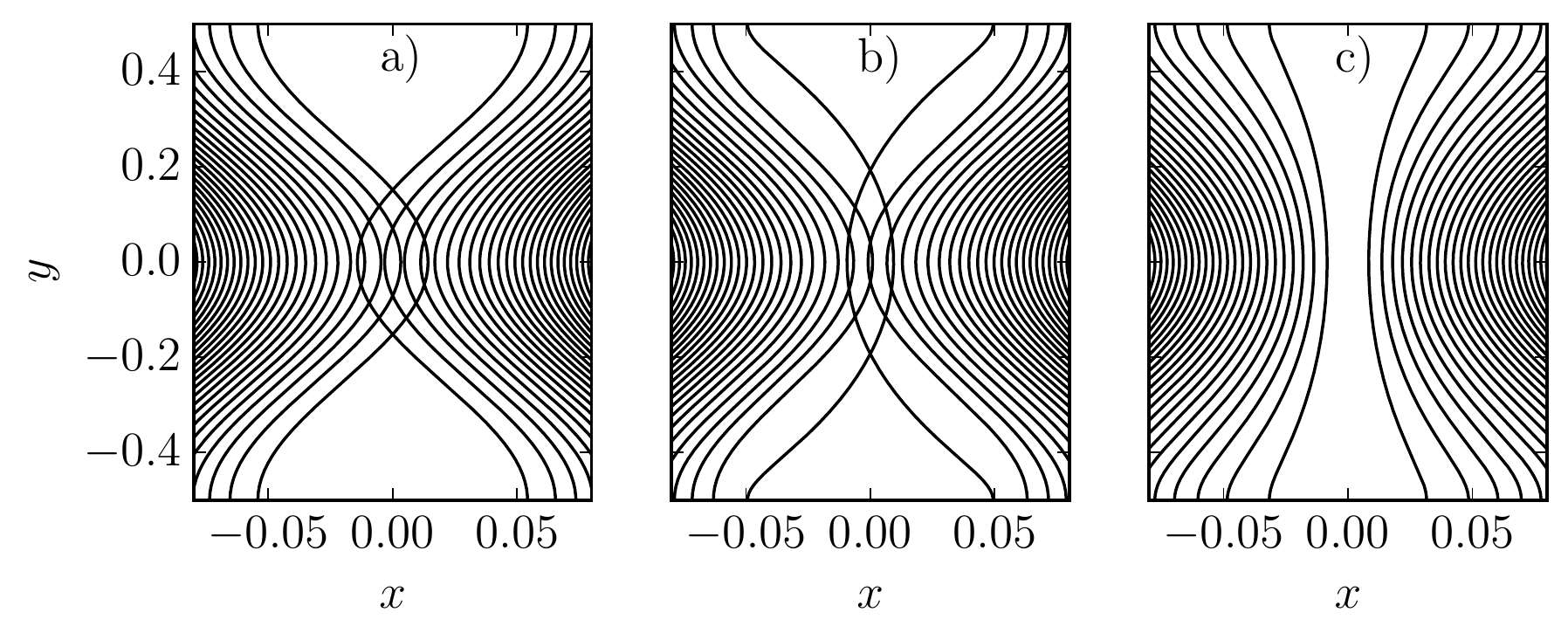}
\caption{\label{pathology}Contours of flux function subject to displacement with $\delta=0.05$: 2D solution (\ref{HKS}) (a), and 3D solutions at the mid-plane for $L=64$ (b) and $L=16$ (c). The intersection of contours in (a) and (b) are pathological.}
\end{figure}

In Fig.\,\ref{pathology}(a), it is shown that the flux surfaces (constant surfaces of flux function $\psi_0=x_0^2/2$) overlap when they are subject to the perturbed fluid mapping $\mathbf x =\mathbf x_0+\bm\xi$, with $\bm\xi$ given by the 2D linear solution (\ref{HKS}). The cause for this unphysical overlap is that the mapping $x(x_0,0)$ is non-monotonic: $\partial x/\partial x_0=1+\partial \xi_x/\partial x_0<0$ at $(0,0)$, since $ \xi_x( x_0,0)\sim-\text{sgn}(x_0)$. In this case, $\mathbf x$ becomes pathological because its Jacobian determinant $J$ is no longer everywhere positive.

However, in order for the perturbed mapping $x(x_0)$ to be non-monotonic, it is not necessary that $\partial \xi_x/\partial x_0$ is singular, as in the 2D solution. The requirement is $\partial \xi_x/\partial x_0<-1$, i.e., the gradient of the displacement is steep enough. For the linear solutions $\xi_x(x_0,0,L/2)$ that are obtained in Sec.\,\ref{linearR}, this can be achieved when the system is sufficiently long.

As shown in Fig.\,\ref{pathology}(b), the flux surfaces can indeed overlap at the mid-plane when subject to the perturbed fluid mapping given by the 3D linear solution for a long system. When the system length is below a critical value $L_f$, the pathology is absent and the flux surfaces do not overlap, as is shown in Fig.\,\ref{pathology}(c). Here the specific value of the critical length $L_f\approx29$ depends on the parameters we obtain the linear solutions with, $\delta$ in particular.

It is worthwhile to emphasize that in 2D, $\partial \xi_x/\partial x_0$ is singular, so the pathology exists for infinitesimal amplitude. In 3D line-tied geometry, $\partial \xi_x/\partial x_0$ is smooth, and finite amplitude is required to trigger the pathology at a critical length. 

Recall that $\partial x/\partial x_0=0$, the trigger for this finite-amplitude pathology, is also a signature of the current singularity that is identified in the nonlinear solution to the 2D HKT problem. Interestingly, \citet{Loizu2015b} have also linked similar finite-amplitude pathologies of linear solutions to the existence of current singularities in 3D equilibria. We therefore suspect that the nonlinear solution to the line-tied HKT problem may be singular above a finite length, which is presumably comparable to the critical length $L_f$ for the finite-amplitude pathology of the linear solution. We investigate whether our nonlinear results support this speculation in Sec.\,\ref{nonlinear}.

 \subsection{Nonlinear results}\label{nonlinear}
We solve the line-tied HKT problem numerically using the method described in Sec.\,\ref{numerical}, in a domain of $[-a,a]\times[-\pi/k,\pi/k]\times[0,L]$. The perfectly conducting walls at $x_0=\pm a$ are deformed into to the shapes that $x(\pm a, y_0,z_0) =\pm[a-\delta\cos ky(\pm a, y_0,z_0) \sin(\pi z_0/L)]$. The boundary conditions in $y$ and $z$ are periodic and no-slip, respectively. We use the same parameters as used in the linear study, namely $a=0.25$, $k=2\pi$, $\delta=0.05$, with varying $L$.

{In practice, we find that full MHD simulations of this problem often exhibit unphysical motions in the $z$ direction, which are numerically unstable. To avoid this issue,} we adopt the RMHD approximations described in Sec.\,\ref{RMHD}, by setting $B_{0z}=1$ and $z=z_0$. For the sake of numerical practicality, we use moderate pressure to approximate incompressibility, instead of enforcing the constraint $J=1$. After all, incompressibility itself is an approximation. Specifically, we initialize with $p_0=0.1-x_0^2/2$ to balance the sheared field $B_{0y}=x_0$, and choose $\gamma=5/3$. In our numerical solutions, we find $|J-1|\lesssim 0.02$, which decreases as resolution increases. This suggests that our solutions are very close to incompressible.

A consequence of approximating $J=1$ is that Eq.\,(\ref{reduced}) does not hold exactly anymore, but instead we have $\mathbf{B}\cdot\nabla j_z=\mathbf{j}\cdot\nabla B_z$ with $B_z=1/J$, which is approximately constant. Since the system is symmetric under rotation by $\pi$ with respect to the $z$ axis ($x,y=0$, the field line of interest in this problem), one finds that $\mathbf{B}_{\perp}=\mathbf{j}_{\perp}=0$, and therefore $j_z(z)=j_z(0)/J(z)$, along the $z$ axis. So in our solutions $j_z(0,0,z)$ should still be approximately constant, as will be confirmed by our numerical results.

We use a tetrahedral mesh where the vertices are arranged in a structured manner with resolution $N\times2N\times NL/4$. The grid number in $z$ varies with $L$ so that the grid size does not. The vertices are distributed {uniformly in $z$}, while non-uniformly in $x$ and $y$ to devote more resolution near the $z$ axis. We use a same profile of mesh packing for a given $L$, but adjust it accordingly when $L$ varies.

\begin{figure}[h]
\includegraphics[scale=0.45]{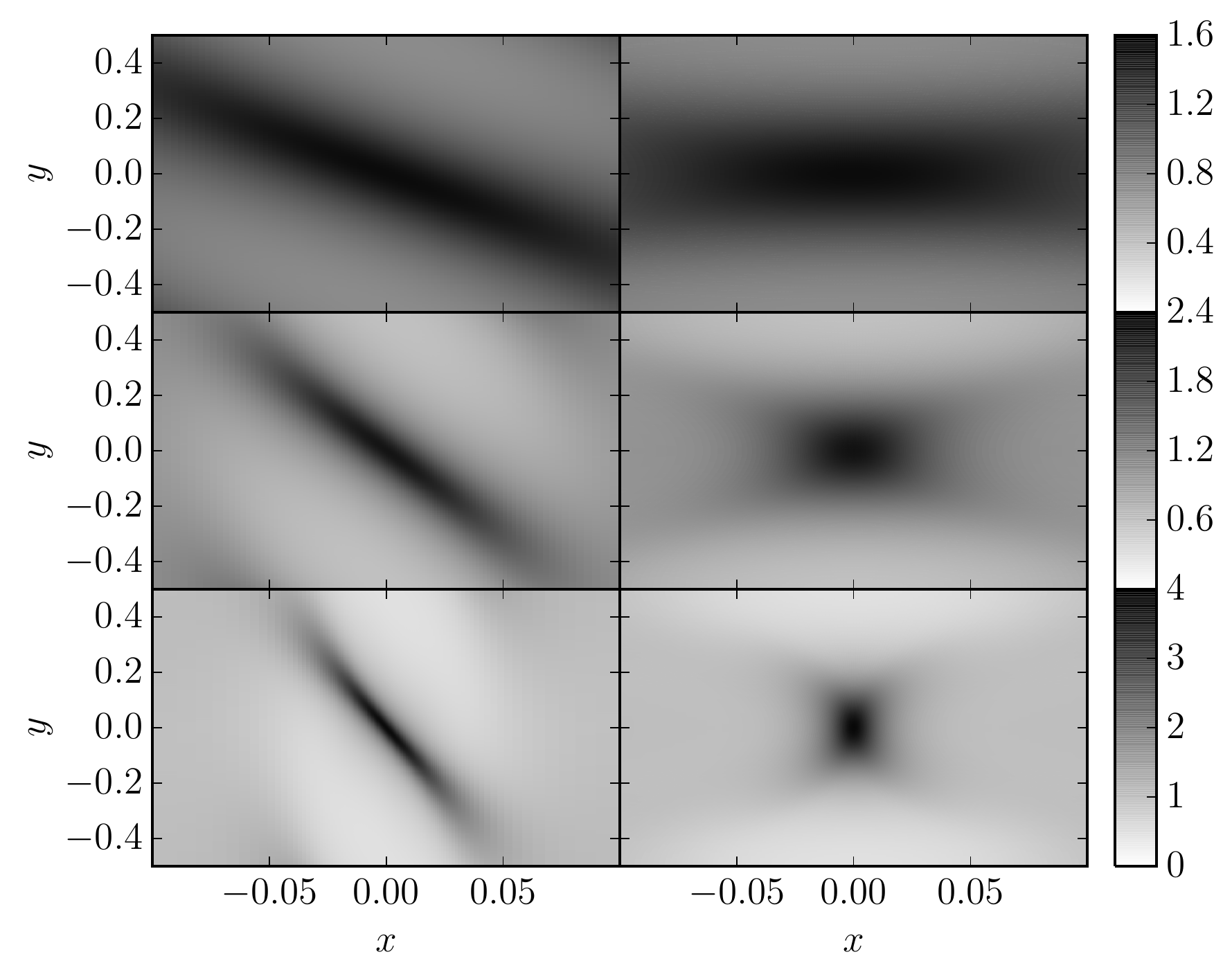}
\caption{\label{current}Distribution of current density $j_z(x,y)$ obtained with $N=160$ for $L=6$, $12$, and $18$ (from top to bottom, respectively) at $z=0$ (left) and $z=L/2$ (right).}
\end{figure}

The system starts from a smoothly perturbed configuration $\mathbf{x}$ consistent with the boundary conditions and relaxes to equilibrium. The specific choice of the initial fluid configuration does not affect the equilibrium solution. In Fig.\,\ref{current}, the equilibrium current density distributions obtained with $N=160$ for $L=6$, $12$, and $18$ are shown. Despite that the distributions become significantly more concentrated to the $z$ axis with increasing $L$, all these solutions turn out to be smooth and well-resolved, as our convergence study shows. 

\begin{figure}[h]
\includegraphics[scale=0.45]{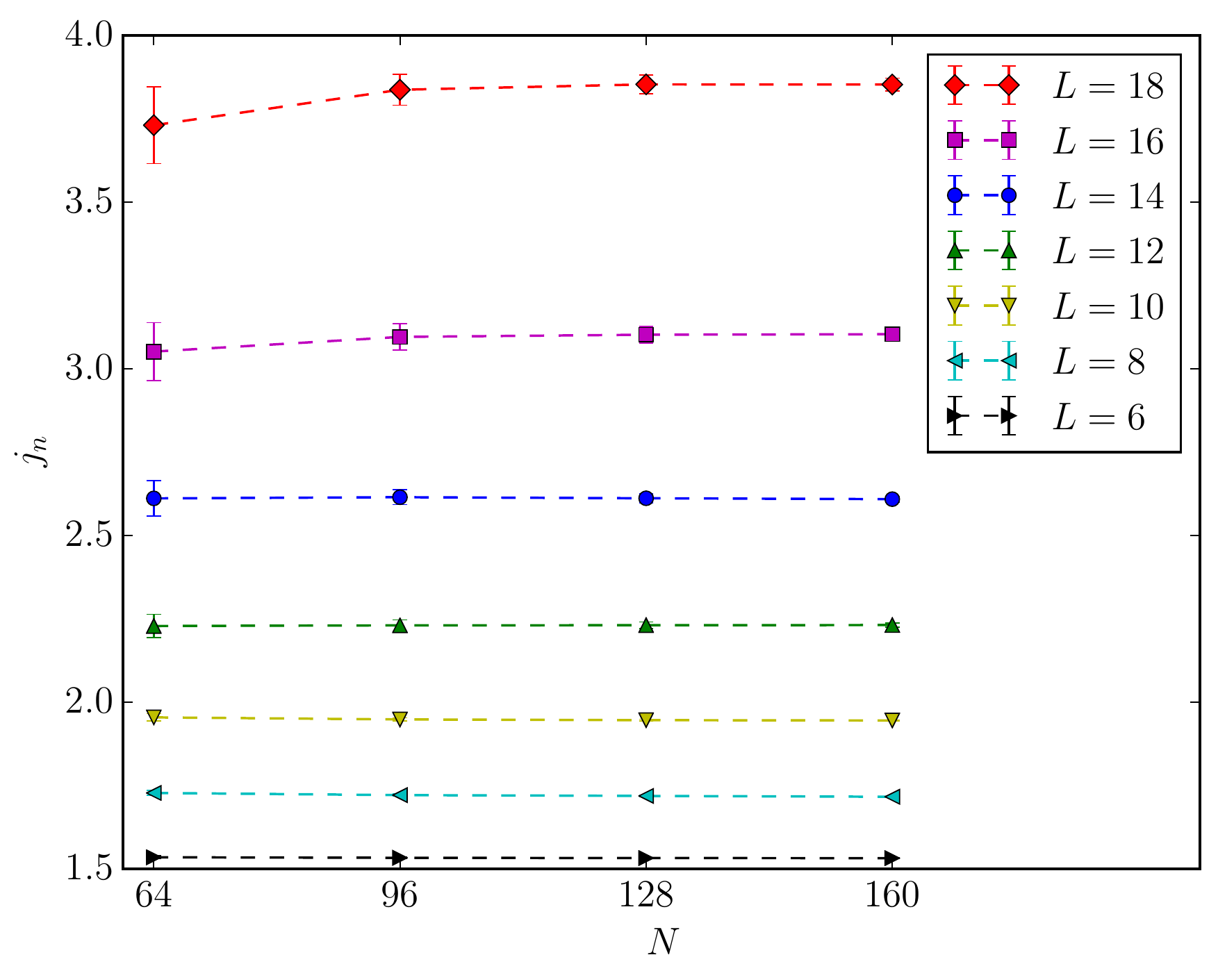}
\caption{\label{resolution}Means ($j_n$) and standard deviations (error bars) of the current density $j_z(0,0,z)$ for varying system length $L$ are shown to converge with increasing resolution $N$.}
\end{figure}

In Fig.\,\ref{resolution}, the means of $j_z(0,0,z)$ (labeled $j_n$) for varying $L$ are shown to converge with increasing resolution $N$. In addition, the standard deviations of $j_z(0,0,z)$, as shown by the error bars, decrease with increasing $N$. That is, $j_z(0,0,z)$ is indeed approximately constant, demonstrating that our almost incompressible solutions capture the features of the exact RMHD solutions very well. We therefore conclude that the nonlinear solutions for these relatively short systems are smooth.

Another observation from Fig.\,\ref{resolution} is that the standard deviation of $j_z(0,0,z)$ increases with $L$. The reason is, for longer systems, the footpoints are more difficult to resolve than the mid-plane. At the footpoints, there is no in-plane motion, which means the mesh there stays as initially prescribed. Meanwhile, as $L$ increases, the mid-plane bears more resemblance with the 2D case, where the squashing effect spontaneously packs the {in-plane mesh towards the $z$ axis. In comparison with the mid-plane, near the footpoints, we need to pack the mesh more aggressively towards the $z$ axis in the simulations, in order to compensate for the self-packing near the mid-plane, particularly for longer systems.} To sum up, longer systems are simply much more challenging to resolve computationally than shorter ones.

%However, things get more complicated as $L$ increases. For $L=20$ and $22$, $j_z(0,0,0)$ appears to have saturated with resolution $N$, while $j_z(0,0,L/2)$ does not. The two values have not agreed, also suggesting that the solutions have not converged yet. Nonetheless, we expect that for these lengths convergence can eventually be achieved, should one keep increasing resolution.

%For $L=24$ and higher, both $j_z(0,0,0)$ and $j_z(0,0,L/2)$ keep increasing with resolution $N$, and their values are far from agreeing.

%Another observation from Fig.\,\ref{resolution} is that with increasing $L$, the discrepancy between $j_z(0,0,0)$ and $j_z(0,0,L/2)$ gets larger. This is better seen in Fig.\,\ref{length}, where $j_z(0,0,0)$ and $j_z(0,0,L/2)$ are now plotted against system length $L$ for varying resolution $N$. For given $N$, $j_z(0,0,0)$ saturates with $L$, which is even weaker than the linear scaling predicted by the linear solution (Fig.\,\ref{deltaj}). This suggests that the footpoint is more under-resolved, in comparison with the mid-plane.

\begin{figure}[h]
\includegraphics[scale=0.45]{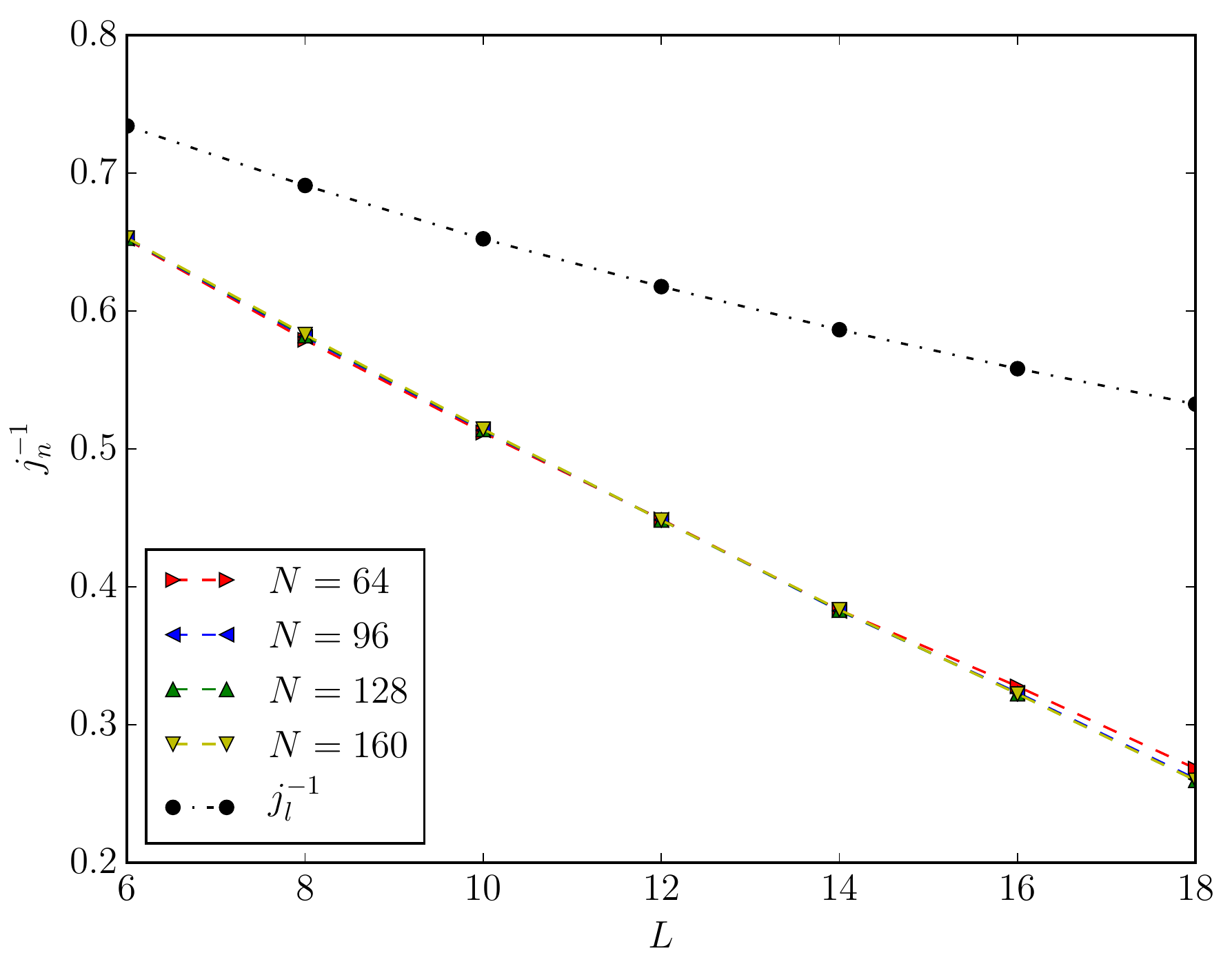}
\caption{\label{length}Inverse of the mean ($j_n$) of the current density $j_z(0,0,z)$ (dashed lines) with varying resolution $N$ vs.\,system length $L$. $j_n\sim (L_n-L)^{-1}$ can roughly be observed. Linearly calculated current density $j_l$ (dashed-dotted line) is also shown for comparison (solutions with $N=128$ from Fig.\,\ref{deltaj} are used).}
\end{figure}

Still, one wonders whether the nonlinear solution is smooth for arbitrary $L$. Fig.\,\ref{length} shows that $j_n^{-1}$ decreases roughly linearly with $L$. That is, $j_n\sim (L_n-L)^{-1}$, which diverges at a critical length $L_n$. This suggests that the nonlinear solution may become singular at finite length. Using the solutions with $N=160$ for fitting, we obtain $L_n\approx25.81$, which is comparable to the critical length $L_f$ for the finite-amplitude pathology discussed in Sec.\,\ref{finite}. Fig.\,\ref{length} also shows that $j_n$ is larger than $j_l$, as expected.

In order to validate such a diverging scaling law and confirm the existence of the finite-length singularity, we should examine the solutions for systems with lengths close to or above the critical value $L_n$. Unfortunately, we are not able to obtain (converged) equilibrium solutions for systems with $L=20$ or higher: as $j_z(0,0,z)$ increases with $L$, the motion near the footpoints becomes more strongly sheared, {distorting the mesh severely, as discussed in Secs.\,\ref{RMHD} and \ref{numerical}. Eventually, $J$ becomes not everywhere positive, and the simulation terminates, however small the time step is}. As $L$ increases, $j_z(0,0,z)$ may indeed diverge at a finite length, or convert to a different scaling law that stays well-defined for arbitrary $L$. With the results in hand, we cannot confirm or rule out either possibility conclusively.

%To sum up, we have shown that the nonlinear solution to the line-tied HKT problem is smooth when the system is short. As it becomes longer, approaching the critical length for the pathology in the linear solution, it appears that the nonlinear solution may become singular above a finite length. Unfortunately,  

\section{Discussion}\label{discussion}
One conclusion we can indeed draw from our results is that 3D line-tied geometry does have a smoothing effect on the current singularity in the 2D HKT problem. In 2D, both the linear and nonlinear solutions are singular. In 3D line-tied geometry, the linear solution is smooth for arbitrary system length; the nonlinear solution is smooth when the system is short. 

Whether the nonlinear solution becomes singular at a finite system length remains yet to be confirmed. Our numerical results show that the maximum current density scales with $(L_n-L)^{-1}$, which implies finite-length singularity. However, since we cannot obtain numerical solutions to validate such a scaling law near the critical value $L_n$, these results can only be considered suggestive, but not conclusive. Nonetheless, we remark that this scaling law is already stronger than the exponential scaling predicted by \citet{Longcope1994b}. To our knowledge, such a scaling law has never been previously reported.

In this paper, we have prescribed what we believe is an effective formula for realizing possible current singularities in 3D line-tied geometry. The idea is to extend a 2D case with singularity to 3D line-tied geometry, and then make the system really long. In particular, the HKT problem is arguably a simplest prototype, for it captures how a sheared field responds to squashing, both ingredients ubiquitous in nature. Also, the finite-amplitude pathology in its linear solution may be suggestive for the possible finite-length singularity in the nonlinear solution.

The results of the HKT problem can also be suggestive for other cases with more complex magnetic topologies, such as the internal kink instability \citep{Rosenbluth1973,Huang2006} and the coalescence instability \citep{Longcope1993,Longcope1994,Longcope1994b}. The obvious distinction between the HKT problem and these cases is the former is externally forced, while the latter are instability driven. A subtlety is, for the instability driven cases, the linear equilibrium equation usually has no nontrivial solutions. In these cases, (fastest-growing) eigenmodes are usually considered to as linear solutions. However, eigenmodes do not have intrinsic amplitudes, unlike in the HKT problem where the linear equilibrium solution can reasonably be given the finite amplitude of the boundary forcing. Consequently, the linear solutions in the instability driven cases can be less suggestive for the nonlinear ones. 

Nonetheless, critical lengths still exist in 3D line-tied geometry for the instability driven cases. That is, these systems are unstable only with lengths above certain finite values \citep{Longcope1994,Longcope1994b,Huang2006}. In fact, \citet{Ng1998} argued that current singularities must emerge when these line-tied systems become unstable. %The instability driven cases will be discussed in more detail in our forthcoming paper on the coalescence instability in 3D line-tied geometry.

Still, what prevents us from obtaining more conclusive results is the limitation of our numerical method, namely its vulnerability to {severe mesh distortion} caused by strongly sheared motion. There are a few remedies that are worth investigating. One option is to enforce incompressibility ($J=1$), since the signature of {severe mesh distortion is $J$ becoming not everywhere positive}. However, naively enforcing this constraint means implicitly solving a global nonlinear equation at every time step, which is not practical. What might be a better option is to solve for pressure from a Poisson's-like equation, yet that could still be expensive on an unstructured mesh. More importantly, when the motion becomes too strongly sheared for the mesh to resolve, enforcing incompressibility may just not be enough.

An alternative is to employ adaptive mesh refinement. Intuitively, that means to divide a simplex into smaller ones once its deformation reaches a certain threshold. This approach will not work for problems with strong background shear flows where the number of simplices can grow exponentially, but may suffice for the Parker problem that is quasi-static. In addition, one may consider more delicate discretization of Eq.\,(\ref{momentum3}) to make the mesh itself more robust against shear flow.

Finally, we emphasize that the Parker problem is still open and of practical relevance, by echoing the latest review by \citet[][p.\,11]{Zweibel2016}: {``It is important to determine whether the equilibrium of line-tied magnetic fields has true current singularities or merely very large and intermittent currents, to characterize the statistical properties of the sheets and to determine how the equilibrium level and spatial and temporal intermittency of energy release depend on $S$."}

\acknowledgments
We thank E.~G.~Zweibel for helpful discussions and comments on an early version of this manuscript. This research was supported by the U.S.~Department of Energy under Contract No.~DE-AC02-09CH11466, and used resources of the Oak Ridge Leadership Computing Facility at the Oak Ridge National Laboratory, which is supported by the Office of Science of the U.S.~Department of Energy under Contract No.~DE-AC05-00OR22725.

\end{document}